\documentclass[conference]{IEEEtran}
\IEEEoverridecommandlockouts

\usepackage{cite}
\usepackage{amsmath,amssymb,amsfonts}
\usepackage{algorithmic}
\usepackage{graphicx}
\usepackage{textcomp}

\usepackage{url,hyperref,lineno,microtype,subcaption}
\usepackage{comment}
\usepackage[dvipsnames]{xcolor}
\usepackage{braket}
\usepackage{soul}
\usepackage{physics}

\usepackage[boxruled]{algorithm2e}
\SetAlgorithmName{Protocol}{protocols}{List of Protocols}
\SetKwInput{kwSetup}{Setup}
\SetKwInput{kwOutput}{Output}

\newcommand{\black}[0]{\color{black}}

\def\BibTeX{{\rm B\kern-.05em{\sc i\kern-.025em b}\kern-.08em
    T\kern-.1667em\lower.7ex\hbox{E}\kern-.125emX}}

\begin{document}

\title{Building Trust in the Quantum Cloud with \\Physical Unclonable Functions}

\author{
\IEEEauthorblockN{Behnam Tonekaboni\IEEEauthorrefmark{1}\IEEEauthorrefmark{4}, Pranav Gokhale\IEEEauthorrefmark{2}, Kaitlin N. Smith \IEEEauthorrefmark{3}}
\IEEEauthorblockA{\IEEEauthorrefmark{1}Infleqtion, Melbourne, VIC, Australia}
\IEEEauthorblockA{\IEEEauthorrefmark{2} Infleqtion, Chicago, IL, USA}
\IEEEauthorblockA{\IEEEauthorrefmark{3}Northwestern University, Evanston IL, USA}
\IEEEauthorblockA{\IEEEauthorrefmark{4}email: behnam.tonekaboni@infleqtion.com
}
}
\maketitle
\begin{abstract}
As cloud-based quantum computing expands, securing access to quantum hardware is increasingly critical. We present an authentication protocol that leverages intrinsic quantum device properties to construct Quantum Physical Unclonable Functions (Q-PUFs). Using frequency fingerprints from fixed-frequency transmon qubits, we prototype our approach on IBM quantum devices with both real and simulated data. We employ fuzzy extractors to generate stable cryptographic keys that tolerate measurement noise and conceal raw hardware data. To support scalability, we introduce q-tuples (qubit subsets) that enable challenge–response generation for ``strong'' PUF behavior. We also outline extensions to neutral-atom platforms and propose future directions including logical Q-PUFs. Our work lays the groundwork for secure, hardware-rooted authentication in hybrid quantum–classical systems.

\end{abstract}

\begin{IEEEkeywords}
Quantum computing, quantum network security, quantum system security, hardware security, physical unclonable functions.
\end{IEEEkeywords}


\section{Introduction}
\subsection{The need for Quantum Device Authentication}

Emerging quantum computers (QCs) and their corresponding software stacks are primarily accessed by end users via the cloud using internet connectivity and runtime credits. Current cloud vendors operating their own QCs include industry giants like Google and IBM, as well as startups such as Infleqtion, Quantinuum, IonQ, Rigetti, and Xanadu. Furthermore, Amazon Braket and Microsoft Azure Quantum offer quantum computing as a service through various quantum hardware providers.

Cloud-based quantum computing has evolved from a niche interest to a topic of mainstream attention in the scientific community over the past five years. Over the next decade, it is conceivable that remotely accessed networks comprising QCs and their associated classical control systems will be adopted at an accelerated rate by scientists across academic, industrial, and defense sectors. As new quantum applications continue to emerge, these hybrid quantum/classical systems will be applied to sensitive tasks with significant social and economic implications, including breaking cryptographic systems, discovering improved materials, exploring complex biological and chemical systems, and optimizing infrastructure for autonomous and robotic systems.

\begin{figure}[h]
     \centering
         \includegraphics[width=\linewidth,trim={0 1cm 0 1cm},clip]{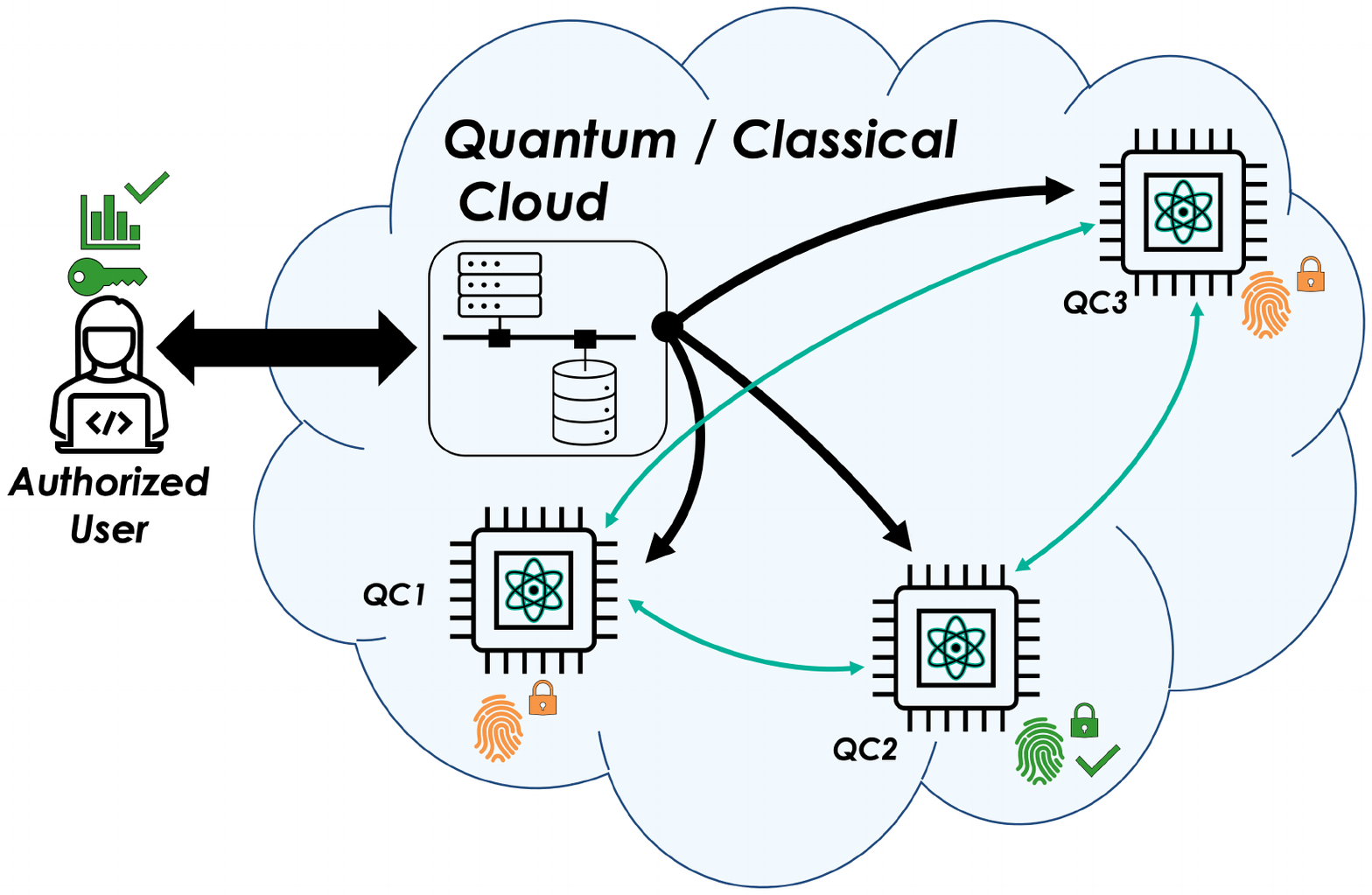}
        \caption{Physical / intrinsic quantum physical unclonable functions (Q-PUFs) provide a means to uniquely identify quantum computers, or QCs, in the hybrid quantum / classical cloud. Physical Q-PUFs rely on robust and reproducible property signatures of the quantum hardware to establish security and trust.
        }
        \label{fig:overview}
\end{figure}

Quantum-enabled systems offer many opportunities to solve problems beyond the capabilities of classical machines. However, these emerging quantum systems may also face novel security challenges. Building secure and trustworthy hybrid quantum-classical networks is essential for the future. It is crucial to identify potential security gaps and vulnerabilities that could arise in high-performance computing (HPC) systems as quantum hardware is integrated. In response to this need, we aim to develop techniques for identifying and authenticating quantum resources within hybrid computing environments.

To prevent disruption to QCs and ensure their reliability and resilience, we propose an approach for fingerprinting and utilizing Quantum Physical Unclonable Functions (Q-PUFs) within quantum systems. Our protocol draws inspiration from the classical concept of Physical Unclonable Functions (PUFs)~\cite{ruhrmair2014pufs}, leveraging their core idea—harnessing manufacturing-induced randomness for unique and unclonable identifiers—to design a quantum analog suitable for secure authentication in quantum computing systems. An overview of our proposal is shown in Fig.~\ref{fig:overview}. A key component of our authentication protocol is the quantum hardware fingerprint. Hardware fingerprints are a well-known security primitive in classical hardware~\cite{holcomb2007initial, layman2004electronic, FPGA-Fingerprinting}. A hardware fingerprint consists of intrinsic information collected from a remote computing device and is used for hardware identification. A robust fingerprint includes features that are unique to a device and must remain stable over time. Additionally, methods for device identification should be collision-resistant, ensuring minimal correlation between fingerprints from different devices.

Unlike fingerprinting classical computers that are remotely accessed, identifying device-level features that are distinct across different QCs and reasonably stable remains a challenge.  
Classical PUFs are valued for lightweight security in devices. This contrasts with Q-PUFs, which serve a different purpose: authenticating and certifying expensive, cloud-based (shared) quantum hardware. This enables trust in the provenance of quantum computations and lays the foundation for protocols that certify communication with genuine quantum processors.  
Furthermore, it is desirable to incorporate these intrinsic device features into a physically unclonable function (PUF), so that the fingerprint data can be confidently used for cryptographic tasks, where the following are desired: 1) many unique keys, and 2) obfuscation of the true hardware fingerprint.

\subsection{Fingerprinting Quantum Devices}
As QCs are expected to remain cloud-accessed, fingerprinting becomes a critical security measure. Unique device identifiers enable authentication and ensure users are interacting with the correct quantum hardware.

QCs exhibit intrinsic properties---such as qubit connectivity, operation frequencies, coherence times, gate fidelities, and measurement errors---that are unique and can serve as the basis for fingerprinting in a Q-PUF. However, it is vital that a fingerprint comprise of QC properties that exhibit sufficient consistency over time. 

\begin{figure}[t]
     \centering
         \includegraphics[width=0.99\linewidth,trim={0cm 0cm 0cm 0cm},clip]{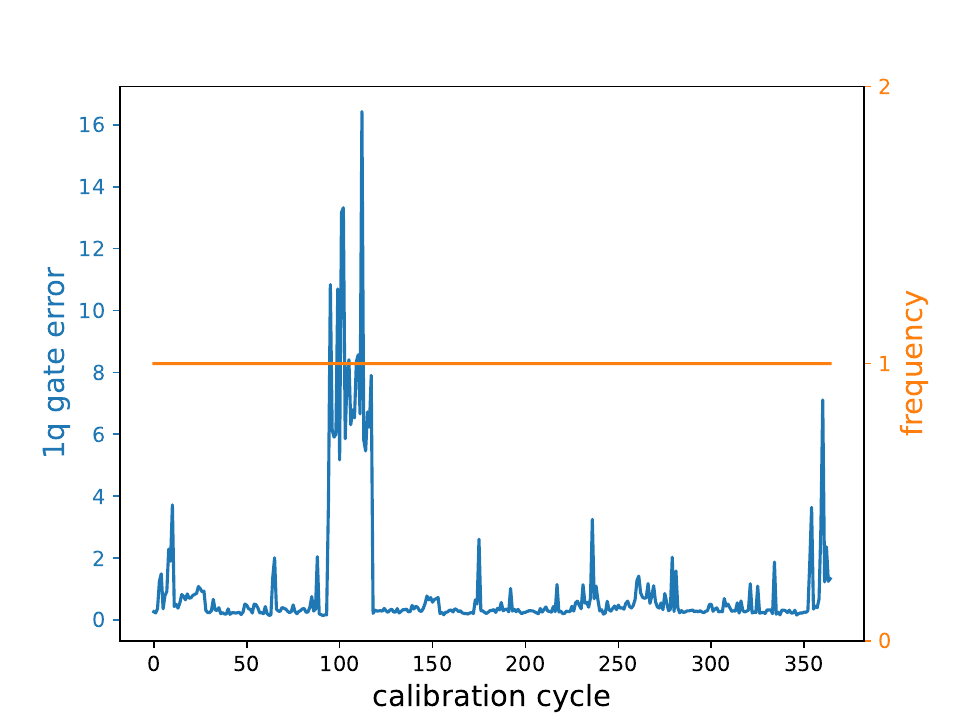}
        \caption{Example of features that are unsuitable and suitable for a Q-PUF basis: single-qubit gate error (left, unsuitable) and qubit frequency (right, suitable) across 365 calibration records (normalized by their respective mean) for IBM Kyiv qubit zero. IBM Kyiv is a 127-qubit QC and was accessed in December 2024.
        }
        \label{fig:device-1q-error-freq}
\end{figure}

Each qubit in a QC carries a range of historical signatures, and when considered collectively, these provide a more holistic view of the machine's individuality. With these insights, elements of a QCs operational signature that are \emph{reasonably} time-invariant can be distilled and leveraged as intrinsic properties for hardware fingerprinting. 
As an example of a property signature, consider the normalized gate-error and frequency plot for a single transmon qubit (qubit $0$ on IBM-Kyiv) shown in Fig.~\ref{fig:device-1q-error-freq}, plotted over $365$ calibration cycles. While the gate-error fluctuates considerably in the plot, the qubit's frequency remains relatively stable and reliable over time, making it a suitable candidate for use in fingerprinting.

\subsection{Contributions} 
We aim to establish robust security guarantees in networks incorporating quantum technologies by proposing methods to prevent attacks where an adversary impersonates a target quantum computer selected for a specific quantum algorithm. The contributions of this work are summarized as follows:

\begin{itemize}
    \item We introduce a Q-PUF based on intrinsic QC fingerprints and use fixed-frequency transmon qubits as our prototype technology.
    \item We describe how fuzzy extraction can be applied to reliably generate large volumes of consistent cryptographic keys using intrinsic QC fingerprints.
    \item We quantitatively evaluate the effectiveness of our Q-PUF through an analysis based on Hamming distance, demonstrating its potential for future validation mechanisms in quantum cloud environments. 
    \item We demonstrate the scalability of our protocol by expanding it to larger devices.
\end{itemize}  

Our goal is to establish foundational tools for detecting and preventing malicious access to, and interference with, quantum computation---thereby ensuring the integrity of quantum systems and their computations.

\section{Fundamentals \& Assumptions}

\subsection{Physical Unclonable Functions (PUFs)}

Strong cryptographic protocols for authentication and communication within classical networks require clearly defined setup assumptions and robust security proofs. Trusted authorities are often employed to issue unique identifiers, such as digital certificates, enabling secure communication between endpoints. However, these trusted authorities can also introduce a significant point of vulnerability within the communication scheme. As a solution, proposals have emerged that use a device’s physical characteristics as a unique fingerprint~\cite{suh2007physical}. Using a physical fingerprint as the basis for identification eliminates many of the overheads associated with trusting a designated party during the setup of a cryptographic scheme, as secret keys do not need to be established by or transferred from an external entity. Hardware-based fingerprints serve as the underlying mechanism that enables physical unclonable functions.

A Physical Unclonable Function (PUF) is a hardware security primitive that leverages intrinsic physical differences between devices. These variations typically arise during the fabrication process, where factors such as manufacturing precision and environmental contaminants introduce random, uncontrollable imperfections. As a result, each device exhibits a unique fingerprint, determined by the distribution and characteristics of these physical defects. Notably, it is nearly impossible for two devices to exhibit identical defect patterns, making PUFs highly effective for secure identification and authentication. \cite{ruhrmair2014pufs}

The random distribution of defects within a device can affect its functionality, resulting in operational properties that are difficult to fully characterize or replicate. For example, random variations in an integrated circuit (IC) can influence the speed at which electrical signals propagate through the circuit’s wires. Standard device operation attempts to mask these random delay characteristics by carefully selecting clock speeds to prevent race conditions, ensuring consistent behavior across devices.
However, PUFs aim to amplify intrinsic differences for the purpose of device identification, using special circuits such as the arbiter PUF and the ring oscillator PUF to exploit unique delay properties between ICs for the purpose of generating cryptographic secrets~\cite{suh2007physical}. Other examples of PUFs used for classical compute hardware include SRAM random initialization~\cite{guajardo2007fpga}, the decay of DRAM modules~\cite{tian2020fingerprinting}, variation in measured resistance~\cite{Helinski2009_resistance}, and optical interference patterns that result when light travels through a waveguide with irregularities~\cite{mesaritakis2018physical}.

PUFs leverage an intrinsic physical fingerprint within a one-way function—a function that is easy to compute given an input but difficult to invert—to establish a challenge-response architecture. When presented with a challenge, the PUF authenticates a device by generating a unique response. For a PUF to be effective in authentication, it must exhibit device uniqueness, temporal consistency, and collision resistance, ensuring minimal correlation across challenge-response pairs. These key properties guided the development of our proposed intrinsic hardware Q-PUF.

\subsection{Variation in Transmon Qubits}
\label{sec:variation-in-qubits}

Transmons are actively researched within academic and industrial settings as they provide a viable platform for realizing physical qubits~\cite{majer2007coupling} with superconducting technology. Transmons employ Josephson Junctions (JJs) along with other hardware at ultra-cold temperatures to act as mesoscopic-scale artificial atoms with an anharmonic energy spectrum. The qubits of interest, fixed-frequency transmon qubits, have their operational frequency set during fabrication -- qubit frequency does not change during quantum computation, unlike other frequency tunable qubit varieties~\cite{barends2013coherent}.
When ``transmon'' or ``qubit'' are mentioned within this work, we assume fixed-frequency quantum devices. Specifically, the development of our Q-PUF focused on prototyping on the IBM fixed-frequency transmon QCs that are available through the IBM Quantum Cloud. 

Similar to classical ICs, transmon qubits are constructed in layers during the device fabrication process~\cite{nersisyan2019manufacturing,place2021new,vrajitoarea2020strongly}. At a high level, qubit fabrication involves: (a) substrate cleaning and preparation, (b) base metal placement, (c) lithography-based feature definition, (d) JJ definition, creation and placement, and (e) packaging and final test~\cite{nersisyan2019manufacturing, place2021new, vrajitoarea2020strongly}.
The practical realities of process imprecision and environmental contaminants mean that each step in the fabrication process is vulnerable to both stochastic and systematic defects. Slight deviations from design targets during fabrication can affect physical qubit characteristics, and abnormalities or defects in physical qubits lead to property variations observable within and across QCs.

The impact of a physical defect on quantum hardware can range from having a subtle impact on operator fidelity to completely rendering parts of the quantum chip inoperable. 
As described in~\cite{smith2023fast}, defects in transmon quantum computers (QCs) result in dynamic (also known as temporal) and static variations in their operational characteristics. In~\cite{smith2023fast}, it was found that qubit characteristics, such as gate fidelity, coherence times, and measurement error, are highly sensitive to hardware variations and imprecise hardware control mechanisms, often resulting in day-to-day fluctuations in these properties. 

On the other hand, the qubit frequency is fixed by the dimensions and placement of the qubit’s active element, the Josephson junction (JJ), and is inherently unique due to imperfections in the fabrication process. Fig.~\ref{fig:device-1q-error-freq} illustrates the difference between static variation (frequency) and dynamic variation (single-qubit operator error) over 365 calibration records for a single qubit (qubit 0) on IBM Kyiv.  To develop a robust Q-PUF, we selected an intrinsic qubit property with static variation, qubit frequency, as the intrinsic hardware fingerprint used as the Q-PUF basis. 

\begin{figure}[t]
     \centering
         \includegraphics[width=0.99\linewidth,trim={0cm 0cm 0cm 0cm},clip]{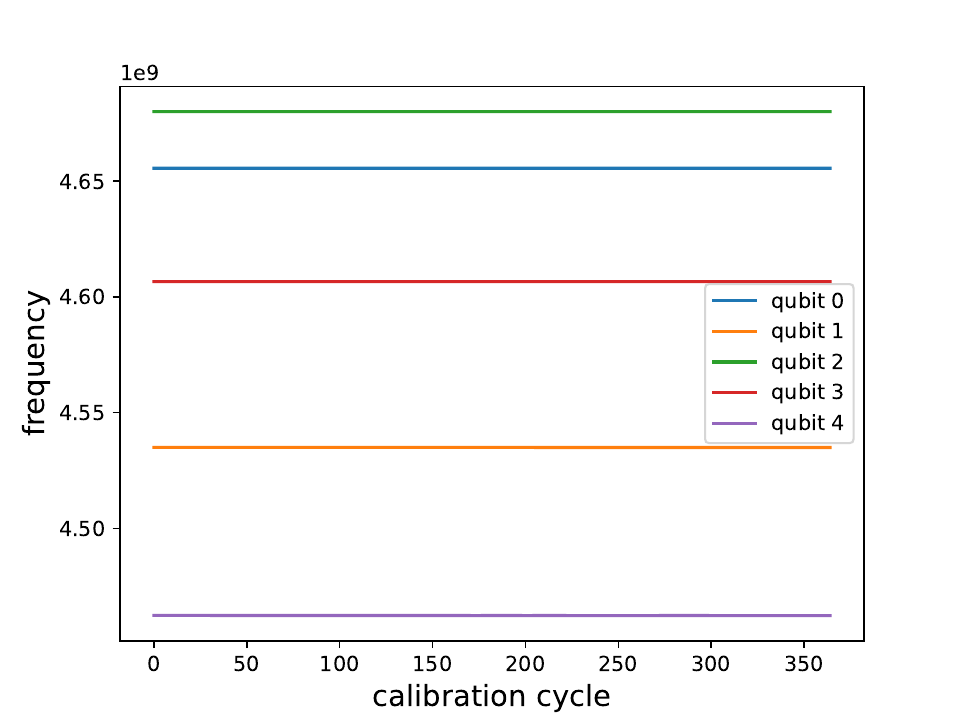}
        \caption{Frequency vs. calibration cycles for five qubits in IBM Kyiv over 365 calibration cycles. This plot shows the consistency of qubit frequency over large timescales. IBM Kyiv is a 127-qubit QC and was accessed in December 2024.
        }
        \label{fig:freq-vs-cycles}
\end{figure}

\section{Superconducting QC Fingerprint for Q-PUF}

\subsection{Qubit Frequencies as Fingerprint}
We focus on remote QCs based on superconducting (SC), fixed-frequency transmon qubits as the subject of authentication via Q-PUF. However, our techniques are extendable to other qubit technologies, as we will discuss later. The QCs in this study are accessed through a network connected to a classical computer. When accessing a QC, the end user aims to uniquely and reliably identify the QC based on its intrinsic, hardware-specific characteristics. Drawing inspiration from~\cite{smith2023fast}, we develop a QC fingerprint for each device, using frequency as its foundation. We assume a secure connection exists between the QC user and the target QC.

In addition to being unique both within and across QCs~\cite{smith2023fast}, qubit frequencies are randomly determined during the fabrication process and exhibit high temporal stability relative to other qubit properties. This stability is illustrated in Fig.~\ref{fig:freq-vs-cycles}, which shows the frequency of five qubits on the IBM Kyiv QC over 365 calibration cycles. Because qubit frequency is set by inherent physical variations during manufacturing, it offers a strong source of entropy and device-specific uniqueness—key requirements for fingerprinting. However, minor variations in control infrastructure precision and environmental conditions can lead to readings that are highly similar but not identical, within a small margin of error. To address this, we found it essential to apply key generation techniques that tolerate low noise thresholds, ensuring robustness under realistic operational conditions for our Q-PUF.

We note that methods exist for adjusting the frequency of fixed-frequency transmons. One such technique, known as laser annealing, selectively tunes qubit frequencies to bring them within a range that improves two-qubit interactions between nearest neighbors~\cite{hertzberg2021laser, zhang2020high}. Laser annealing enhances two-qubit gate fidelity and has been used to improve the yield of fabricated IBM QCs.

While procedures like laser annealing enable device-level improvements by modifying critical fingerprint components, it is important to recognize that altering an intrinsic hardware fingerprint also changes the Q-PUF’s corresponding challenge-response behavior. If a Q-PUF application relies on storing challenge-response pairs in a secure location, such changes require re-enrollment of the device. In cases where qubit frequencies are altered due to QC maintenance, the QC must be considered an updated version of the original device, now possessing a “new” intrinsic fingerprint that incorporates these modifications.

It is worth highlighting that the laser annealing process described in~\cite{hertzberg2021laser} is imperfect and does not achieve target frequencies with 100\% precision. As a result, reproducing two identical QCs via laser annealing remains highly improbable. Therefore, despite the possibility of tuning, qubit frequency continues to serve as a viable and distinctive intrinsic hardware fingerprint.

\subsection{ Characterizing Real QC Devices}
\begin{figure}[t]
     \centering
         \includegraphics[width=0.95\linewidth,trim={.2cm 0cm 0cm 0cm},clip]{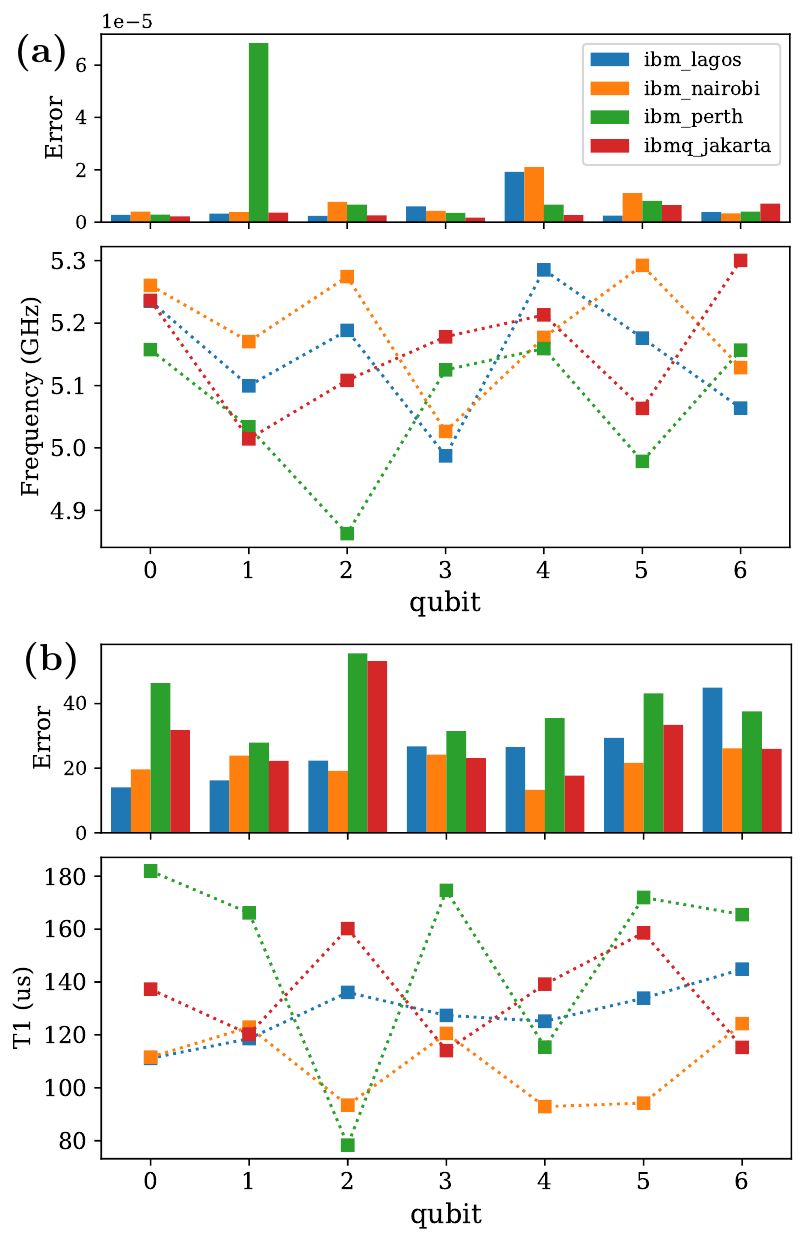}
        \caption{ Examples of the physical properties of qubits across four different IBM devices, along with their corresponding errors. (a) shows the average qubit frequency, while (b) presents the thermal relaxation time ($T_1$). The averages are calculated from $30$ calibration cycles ($\sim30$ days). In each figure, the bottom panel displays the data, and the top panel illustrates the associated errors with bar plots. Different devices are represented by distinct colors. Color legend is true for both (a) and (b). 
        }
        \label{fig:7q-T1-freq}
\end{figure}
The Josephson junction (JJ) is a critical component within a transmon circuit that helps to physically encode qubit state, and the operating frequency of a fixed-frequency transmon qubit is determined by the critical current and capacitance of the transmon's JJ. These parameters are established during fabrication and depend on the physical dimensions and material properties of the active element~\cite{morvan2022optimizing}.  Qubit frequency describes the transition between the qubit's $\Ket{1}$ and $\ket{0}$ state, $f_i = f_{i,01} = f_{\ket{1}}-f_{\ket{0}}$. 
Today's state-of-the-art transmon devices target $f_i$ values in the neighborhood of $\sim 5$ GHz~\cite{zhang2020high}. This is seen on the Fig.~\ref{fig:freq-vs-cycles} plot of frequency data for qubits 0-4 on IBM Kyiv. Additionally, qubit frequency data of four IBM seven-qubit QCs is featured in Fig.~\ref{fig:7q-T1-freq}(a).

The practical implementation of our Q-PUF requires reliable methods to obtain frequency fingerprint data. A spectroscopy experiment is commonly used to determine the frequency of a transmon qubit. This experiment involves probing the qubit with a range of microwave signals to identify its resonant frequency. In IBM QCs, a frequency sweep experiment written in Qiskit can be used to characterize remote devices. In addition, frequency characterization data are made available through the IBM Quantum API. These data, which are updated approximately once per day, are provided alongside other QC calibration data that describe device operational metrics.

We use characterization data from four, seven-qubit IBM QCs—Perth, Jakarta, Lagos, and Nairobi—collected over 30 calibration cycles. The mean frequencies over these cycles are shown in the bottom panel of Fig.~\ref{fig:7q-T1-freq}(a), with their standard deviations indicated as ``Error" in the top panel. Since the standard deviations are negligible compared to the mean values, this demonstrates the consistency of qubit frequencies across the 30 calibration cycles.
Furthermore, as shown in Fig.~\ref{fig:7q-T1-freq}(a), the frequency profile of each device is distinct, effectively creating unique hardware fingerprints for our proposed Q-PUF.

As a counterexample, thermal relaxation time ($T_1$) is not viable for device identification. As shown in Fig.~\ref{fig:7q-T1-freq}(b), the standard deviation (``Error") is significant, in some cases exceeding $50\%$, making $T_1$ an unsuitable choice for fingerprinting.

\subsection{Simulating Fabrication}

\label{sec:fab-sim}

To address the limited availability of real transmon-based quantum devices, we developed a Monte Carlo (MC) simulator to generate synthetic qubit frequency data and evaluate our frequency-based Q-PUF at scale. Modeled after the framework in~\cite{smith2022scaling}, the simulator includes tunable parameters such as qubit topology (e.g., grid, heavy square, or heavy hexagon), target frequencies, the number of qubits per QC, the number of QCs per batch, and fabrication precision.

\begin{figure}[t]
     \centering
         \includegraphics[width=0.95\linewidth,trim={0cm 0cm 0cm 0cm},clip]{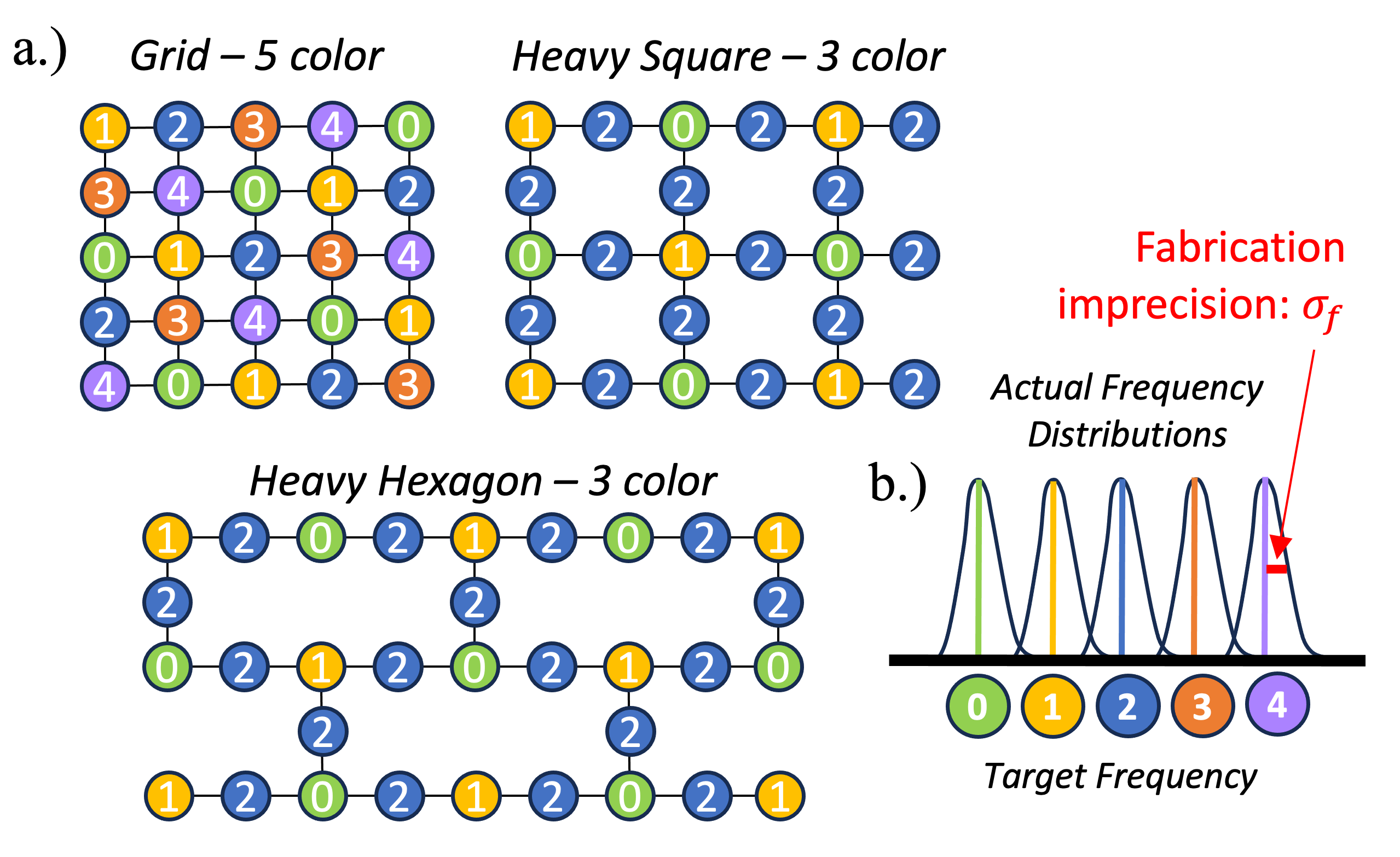}
        \caption{a.) QC topologies modeled by our fabrication simulator. The graphs are colored to show ideal frequency assignment to avoid frequency collisions and maximize fidelity of qubit-qubit entanglement. b.) Ideal, targeted frequencies vs. actual distributions with shapes set by fabrication imprecision, $\sigma_f$.
        }
        \label{fig:topology-freq-imp}
\end{figure}

As described in Section~\ref{sec:variation-in-qubits}, imprecision is an unavoidable challenge during superconducting QC fabrication. As our fingerprinting work is based on frequency, we focus on modeling the defects that deviate a qubit's frequency from its ideal, targeted value within our simulation framework. 
We note that ideal frequency values are chosen according to frequency collision criteria outlined in~\cite{hertzberg2021laser} to maximize the theoretical fidelity of the two-qubit operations on-chip. In~\cite{smith2022scaling}, a step size of 0.06 GHz was found to be optimum in maximizing QC yields (i.e. $f_0 = 5.0$ GHz, $f_1 = 5.06$ GHz, $f_2 = 5.12$ GHz...), so this ideal frequency interval was chosen as default within our MC transmon QC fabrication simulator. The number of target frequencies depends on the selected machine topology since machine graph frequency patterning must avoid frequency collisions. Frequency collisions in adjacent qubits degrades their two-qubit operation fidelity and thus their ability to become entangled~\cite{morvan2022optimizing}. The grid, heavy square, and heavy hexagon QC layouts, and example graph coloring for ideal frequency assignment, are in Fig.~\ref{fig:topology-freq-imp}(a).

Actual qubit frequency variation is stochastic, causing the final frequency profile of each chip to be unique. 
The distribution that describes fabrication precision is characterized by the standard deviation $\sigma_f$, where a lower $\sigma_f$ indicates higher precision and actual device frequencies land near design targets. Ideal, targeted qubit frequencies vs. actual Gaussian distributions with shapes set by $\sigma_f$ is pictured in Fig.~\ref{fig:topology-freq-imp}(b). 
A $\sigma_f = 0.014$ GHz is reported as state of art in~\cite{hertzberg2021laser}. 

\begin{figure}[t]
     \centering
         \includegraphics[width=0.95\linewidth,trim={0cm 0cm 0cm 0cm},clip]{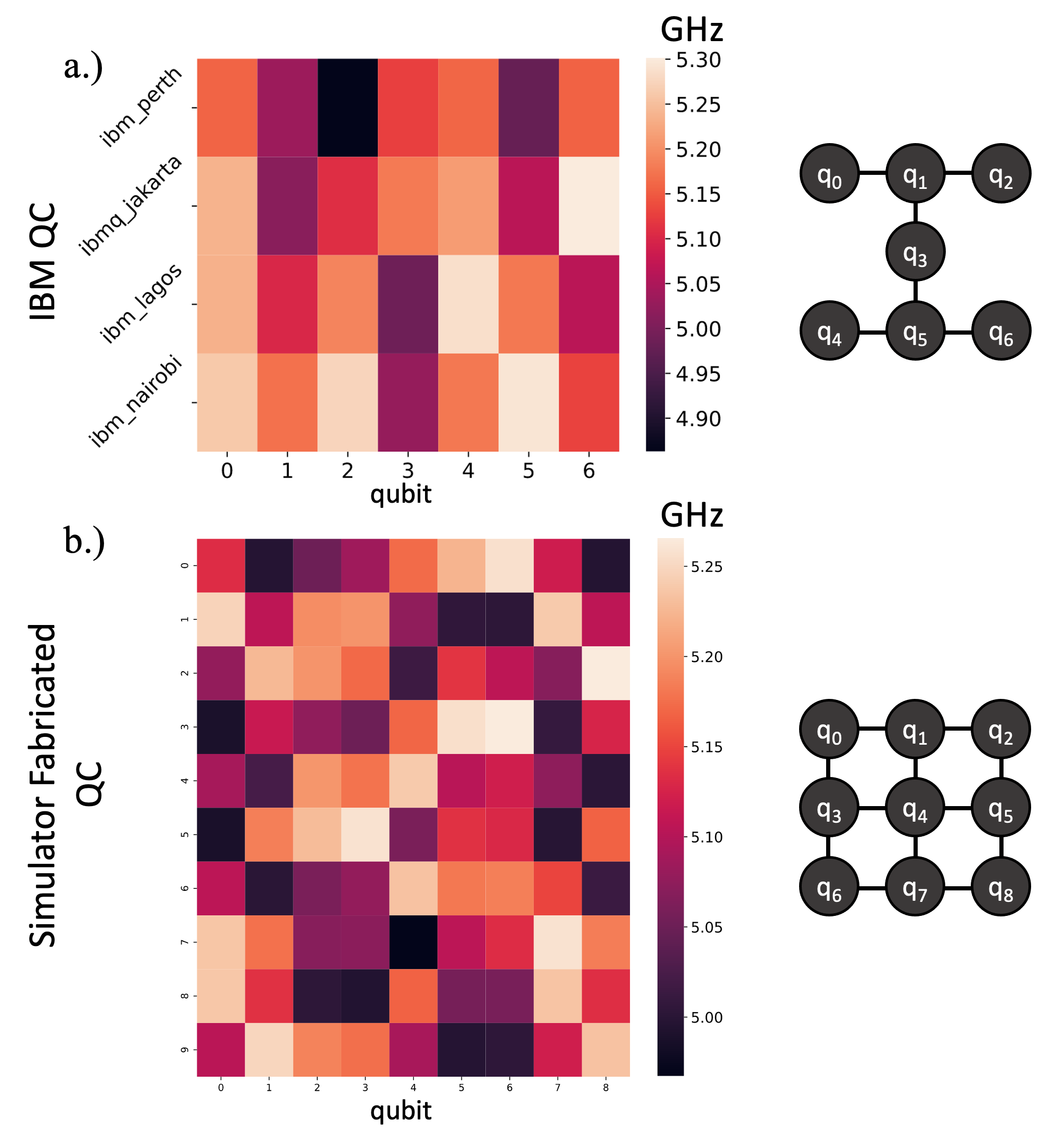}
        \caption{a.) Heat map (left) comparing qubit frequencies across four, seven-qubit IBM QCs and b.) heat map comparing qubit frequencies (left) across ten, nine-qubit QCs generated by our fabrication simulator. In both a.) and b.), QC topology is pictured on the right.
        }
        \label{fig:real-machine-sim-compare}
\end{figure}

Fig.~\ref{fig:real-machine-sim-compare} shows that our fabrication simulator produces qubit frequency distributions that mirror those in today's QCs. Fig.~\ref{fig:real-machine-sim-compare}(a) shows a heat map (left) that compares individual qubit frequencies gathered on the same day in August 2023 for four IBM QCs: Perth, Jakarta, Lagos, and Nairobi (topology pictured on the right). We then simulate the fabrication of ten, nine-qubit QCs, topology shown to the right of Fig.~\ref{fig:real-machine-sim-compare}(b), with our simulator. The heat map to the left of Fig.~\ref{fig:real-machine-sim-compare}(b) compares individual qubit frequencies on the fabricated QCs. This heat map demonstrates that simulated frequencies contain similar randomness as real qubit frequencies. 

\section{ Prototyping Q-PUF on IBM Devices}
\label{tech-methods}

In this section, we prototype a Q-PUF implementation using publicly-accessible IBM superconducting devices. We present the technical methodology behind our prototype, including key extraction via fuzzy extractors and the evaluation of key quality through Hamming analysis. Our goal is to assess whether intrinsic, device-specific quantum properties—such as qubit frequencies—can be leveraged to generate cryptographic keys that are both stable over time and unique across devices.

\subsection{Q-PUF Generation with Fuzzy Extractor}
To generate consistent and sufficiently random keys from the frequency data of a QC, an extractor function is needed after obtaining its intrinsic fingerprint. For this purpose, we use a fuzzy extractor~\cite{dodis2004fuzzy}, a tool designed to transform noisy samples into cryptographically secure keys. Fuzzy extractors are particularly well-suited for this task because they can handle slight variations in fingerprint readings over time. This adaptability makes them a common choice for biometric profiling, where data is often collected from diverse sensors with varying levels of precision.

Our Q-PUF solution utilizes a fuzzy extractor to produce robust keys from the QC frequency fingerprint. As long as the intrinsic QC data is difficult to predict before measurement and remains consistent within an acceptable range across repeated measurements, it can be reliably used to generate cryptographic keys. \black As an added benefit, the fuzzy extractor obfuscates the QCs intrinsic fingerprint data as qubit frequencies assist in producing keys but are not keys themselves. Pseudocode for our prototype that extracts frequency-based Q-PUF keys is shown in Protocol~\ref{alg:qpuf-key-extraction}. 

\begin{algorithm}[t]
\SetKw{KwTo}{to}
$F \gets $ QC qubit frequencies\;
$m \gets $ mean ideal frequency\;
$s \gets $ frequency resolution\;
$e \gets $ error tolerance\;
$id \gets$ [] \;
\For{$f_i$ in $F$}{
$id$.\texttt{append}(\texttt{fmt\_freq\_base\_16}($f_i-m$, $s$))
}
$key$ , $helper\gets$ \texttt{fuzzy\_extract}($id$, $e$)

\caption{Frequency-based Q-PUF key extraction}
\label{alg:qpuf-key-extraction}
\end{algorithm}

The algorithm requires the following inputs: an array of real QC qubit frequencies, $F$; a mean ideal frequency (i.e., the target frequency during fabrication), $m$; a frequency resolution $s$, which sets the precision in terms of significant figures; and an error tolerance $e$ for the fuzzy extractor. To begin, the values in $F$ are formatted. This involves subtracting the constant $m$ from each $f_i$ in $F$, rounding the result to a specified resolution (i.e., a set number of significant digits), and then converting the value into hexadecimal. These new, uniform representations for individual qubit frequencies are then concatenated to form $id$, the intrinsic fingerprint. Next, the $id$ is passed to \texttt{fuzzy\_extract}, implemented using the methods of~\cite{canetti2021reusable}, along with an error tolerance, $e$, which specifies the number of bit-flips the extractor can tolerate. The Q-PUF key extraction procedure outputs a $key$ and a $helper$. The $key$ is the frequency-generated secret used for device authentication and other cryptography protocols. The $helper$, on the other hand, is not secret---it can be public. The $helper$ is produced by the fuzzy extractor and  essential for reproducing the $key$ during later characterizations of qubit frequencies. As long as the frequency values remain within the tolerance threshold $s$, the original key can be reliably reconstructed using the public $helper$. 

The Q-PUF key extraction algorithm plays a central role in enabling secure, hardware-intrinsic key generation and regeneration. To clarify its practical application, we describe a typical usage pipeline that outlines how the system operates from initial enrollment to secure key recovery. This process ensures that the same secret key can be reliably reproduced under noisy conditions without ever storing the key directly.

\subsubsection*{Q-PUF Usage Pipeline}

\begin{enumerate}
    \item \textbf{Enrollment (Initial Characterization):} \\
    Measure the device's QC qubit frequencies $F$, and input them—along with parameters $m$, $s$, and $e$—into the key extraction protocol (Protocol~\ref{alg:qpuf-key-extraction}). This yields a secret $key$ (stored securely) and a public $helper$ (published or stored with the device).

    \item \textbf{Key Reconstruction (Subsequent Characterizations):} \\
    During later use, remeasure the qubit frequencies $F'$, process them identically to produce a new fingerprint $id'$, and apply the fuzzy extractor with $id'$ and the stored $helper$ to recover the original $key$, assuming the deviations are within the tolerance $e$.

    \item \textbf{Authentication and Cryptographic Use:} \\
    The reconstructed $key$ can then be used for cryptographic tasks such as secure key exchange, device authentication, or hardware binding.
\end{enumerate}

We note that, since the $key$ is derived directly from physical device characteristics and reconstructed using the public $helper$, no long-term secret needs to be stored in non-volatile memory. If reconstruction fails—due to excessive noise or tampering—authentication is denied, preserving the integrity and security of the system.
\black

\subsection{Q-PUF evaluation via Hamming Analysis}
We evaluate the effectiveness of our Q-PUF by assessing the quality of the generated keys. The frequency data of the quantum computer serves as the Q-PUF challenge, while the generated key acts as the response. To quantify key quality, we use two metrics: $(1)$ normalized Hamming weight, which reflects the randomness of individual keys, and $(2)$ Hamming distance between keys, which captures their uniqueness. \black

The normalized Hamming weight evaluates a single key by calculating the proportion of $1$s in its binary representation relative to its total length.
A normalized Hamming weight of $0.5$ indicates an even distribution of $0$s and $1$s within a single key, meaning that either value is equally likely. Conversely, a normalized Hamming weight close to $0$ or $1$ indicates a heavily biased key, where one bit value dominates.

\begin{figure}[t]
     \centering
        \includegraphics[width=0.95\linewidth,trim={0cm 0cm 0cm 0cm},clip]{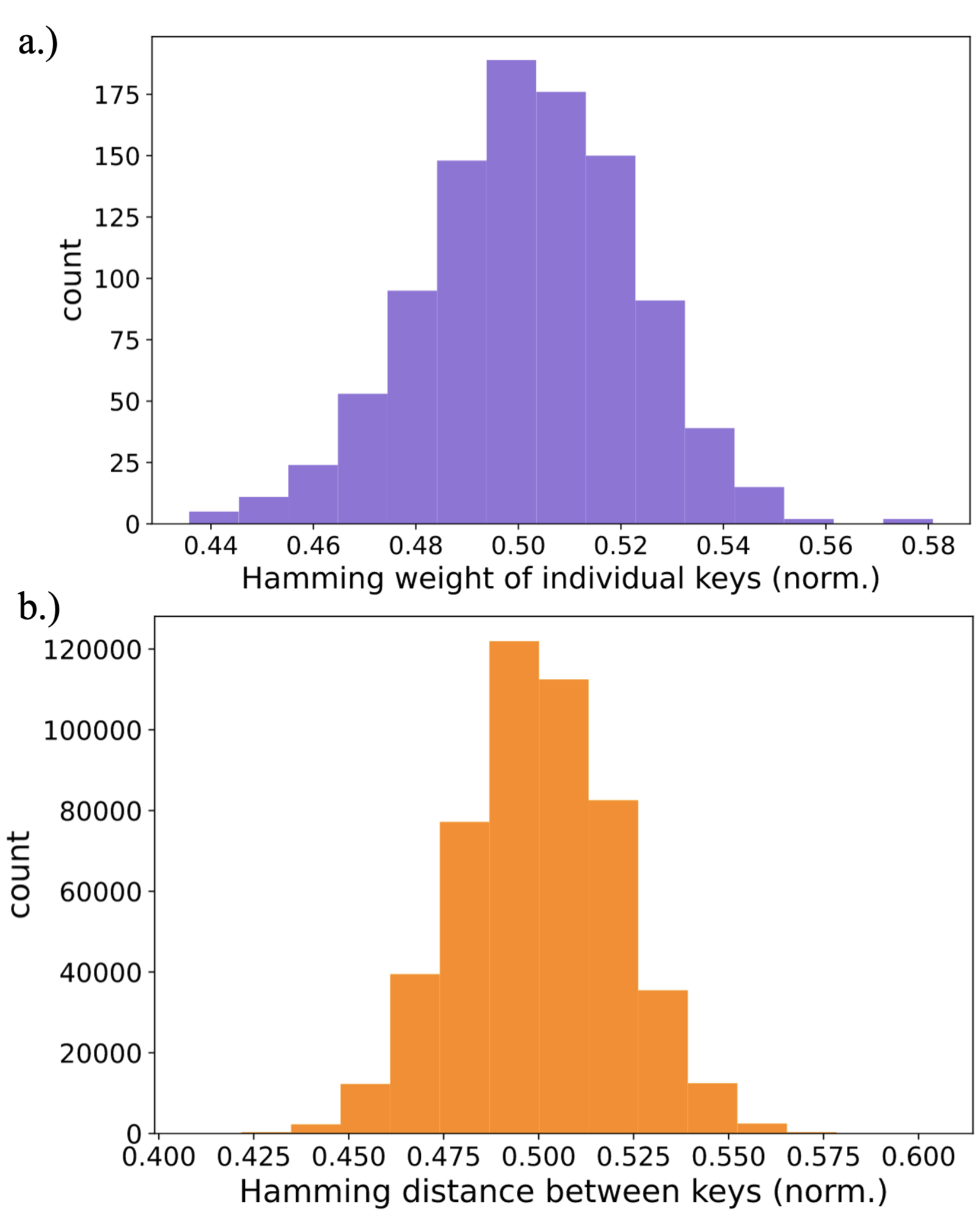}
        \caption{Hamming analysis of Q-PUF keys generated by 1000, nine-qubit simulated QCs. a.) Hamming weight distribution around 0.5, showing the balance of 0s and 1s within keys are sufficiently even. b.) Exhaustive comparison of Hamming distance across all generated keys. A distribution around 0.5 shows little correlation.
        }
        \label{fig:hamming-analysis}
\end{figure}

On the other hand, Hamming distance is used to compare different keys. It calculates the number of bit positions at which two binary values differ, then normalizes this count by the length of the binaries. A normalized Hamming distance of $0$ means the keys are identical, while a value of $1$ indicates that the keys are completely opposite. In cryptographic applications, a mean Hamming distance of $0.5$ is ideal, as it suggests minimal correlation between keys and thus high uniqueness. 

\subsubsection*{Numerical and Experimental Evaluation of Q-PUF Keys}
We evaluated temporal robustness by comparing keys generated across $10$ different calibration cycles of the five-qubit IBM Manila device. Each qubit frequency was encoded using $24$ bits, resulting in a $120$-bit intrinsic fingerprint. During fuzzy extraction, an error tolerance of $8$ bits was applied. Note that, the helper, produced alongside the key in Protocol~\ref{alg:qpuf-key-extraction}, is not used directly in any Hamming distance calculation; rather, it supports error correction by enabling the fuzzy extractor to recover the original key from new frequency measurements that fall within the tolerance threshold $s$. \black The final extracted key for each characterization was $240$ bits in length. An exhaustive pairwise comparison of all generated keys yielded a Hamming distance of zero in every case, indicating perfect consistency across calibration cycles.

To demonstrate the strength of keys generated by our Q-PUF scheme across a large number of QCs, we simulated the fabrication of $1000$ nine-qubit QCs. We collected fingerprint data and applied Protocol~\ref{alg:qpuf-key-extraction} to generate keys. Hamming analysis was then performed, producing the histograms shown in Fig.~\ref{fig:hamming-analysis}. The Hamming weight distribution in Fig.~\ref{fig:hamming-analysis}(a) indicates that the number of $0$s and $1$s in the keys is sufficiently balanced and unbiased---an essential property for cryptographic keys. Furthermore, the Hamming distances in Fig.~\ref{fig:hamming-analysis}(b), obtained from exhaustive pairwise comparisons, show that the keys are highly uncorrelated and random. Notably, no samples appeared near the undesirable extremes of $0$ or $1$ in either histogram.

While Protocol~\ref{alg:qpuf-key-extraction} demonstrates the feasibility of Q-PUF key generation, it remains limited in scale and represents a \emph{weak} PUF. In the next section, we expand this approach to larger quantum devices, highlighting its scalability and showing how it can be extended to realize \emph{strong} PUFs with richer challenge-response behavior.

\section{Expansion and Extension}

\begin{figure}[t]
     \centering
     \includegraphics[width=0.95\linewidth,
         ]
         {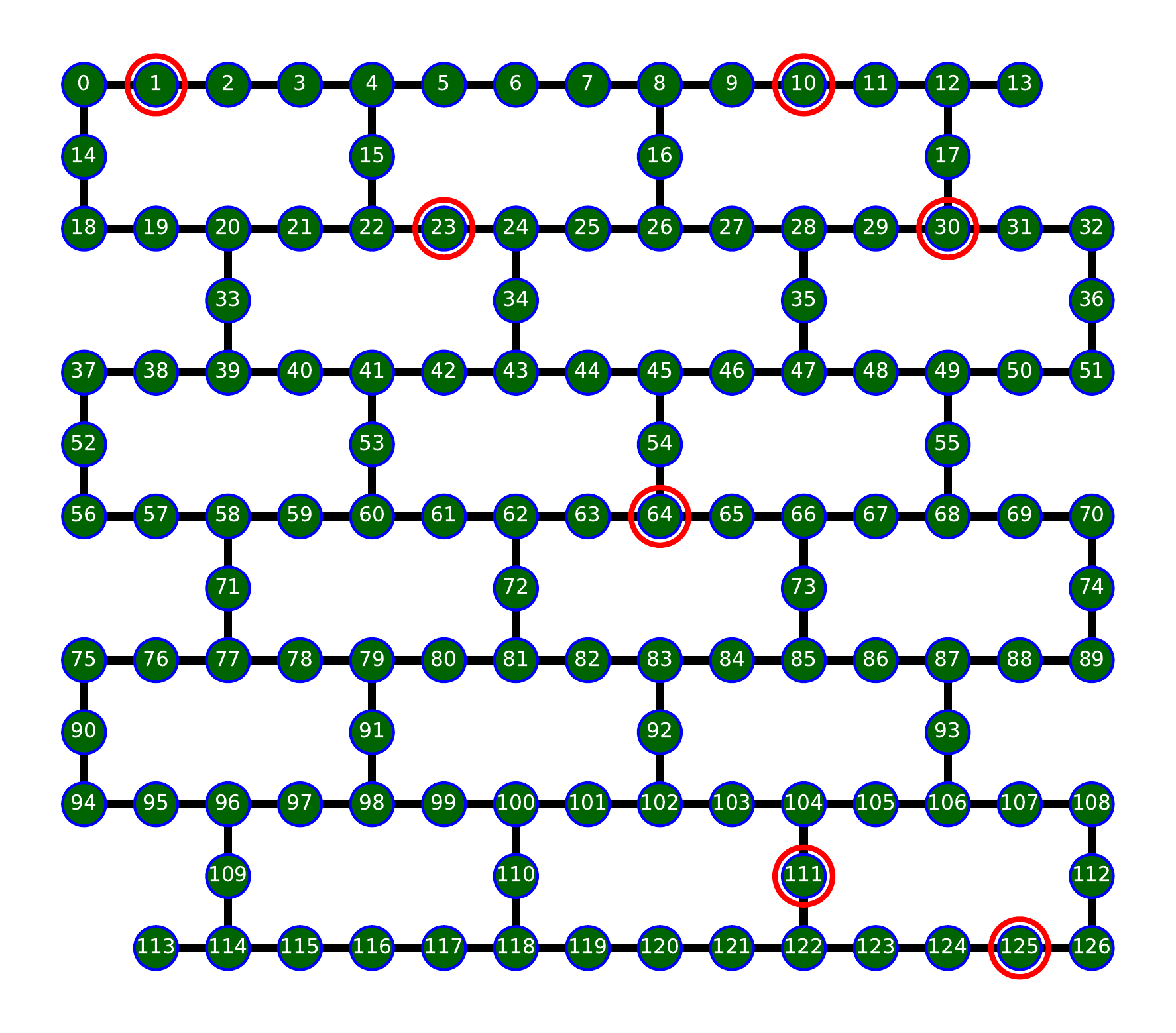}
        \caption{An example of a q-tuple used in the QPUF protocol. The figure displays the topology of IBM's 127-qubit devices, with the qubits forming the q-tuple highlighted by red circles. The frequencies of this 7-qubit subset are sufficient to generate a key for the QPUF protocol.}
        \label{fig:q_tuple}
\end{figure}

In this section, we build on the concepts from Section~\ref{tech-methods} to explain how to implement a practical and scalable Q-PUF. We focus on two key questions: (1) Is the protocol scalable? Specifically, can it be applied to larger devices, or are modifications necessary? (2) Can physical properties other than frequency be used? This would provide additional options for constructing a Q-PUF, beyond relying on a single physical property.

\subsection{Expanding the QPUF method for larger devices}
In the previous section, we introduced our protocol for the Q-PUF and demonstrated its success on IBM devices with a relatively small number of qubits. As the future of quantum computing depends on devices with increasingly large qubit counts, it is crucial that our protocol maintains similar accuracy and reliability on larger systems, as shown earlier. Furthermore, we design our QPUF with practical deployment in mind. To be a valuable cryptographic primitive, the QPUF must do more than generate a single key from intrinsic hardware properties—it must support an architecture capable of producing exponentially many challenge-response pairs at low cost. This aligns with the design goals of a \textit{strong} PUF, \black as outlined in classical literature~\cite{suh2007physical}. The resulting responses must exhibit high entropy and be resistant to collisions. 

\begin{algorithm}[t]
\SetKw{KwTo}{to}
\SetKwFunction{SelectQtuple}{select\_qtuple}
\SetKwFunction{FmtFreq}{fmt\_freq\_base\_16}
\SetKwFunction{FuzzyExtract}{fuzzy\_extract}

$F \gets $ QC qubit frequencies\;
$m \gets $ mean ideal frequency\;
$s \gets $ frequency resolution\;
$e \gets $ error tolerance\;
$k \gets $ q-tuple size\;

$q \gets$ \SelectQtuple{$F$, $k$}
$id \gets$ [] \;

\For{$f_i$ in $q$}{
    $id$.\texttt{append}(\FmtFreq($f_i - m$, $s$))
}
$key$, $helper \gets$ \FuzzyExtract($id$, $e$)

\caption{Weak Q-PUF: key extraction from a fixed q-tuple}
\label{alg:qtuple-qpuf}
\end{algorithm}

\begin{algorithm}[t]
\SetKw{KwTo}{to}
\SetKwFunction{GenerateQtuples}{generate\_qtuples}
\SetKwFunction{FmtFreq}{fmt\_freq\_base\_16}
\SetKwFunction{FuzzyExtract}{fuzzy\_extract}

$F \gets $ QC qubit frequencies\;
$m \gets $ mean ideal frequency\;
$s \gets $ frequency resolution\;
$e \gets $ error tolerance\;
$k \gets $ q-tuple size\;
$CRPs \gets$ [] \;

$\mathcal{Q} \gets$ \GenerateQtuples{$F$, $k$} \tcp*{generate many q-tuples from $F$}

\For{$q$ in $\mathcal{Q}$}{
    $id \gets$ []\;
    \For{$f_i$ in $q$}{
        $id$.\texttt{append}(\FmtFreq($f_i - m$, $s$))
    }
    $key$, $helper \gets$ \FuzzyExtract($id$, $e$)\;
    $CRPs$.\texttt{append}(( $q$, $key$, $helper$ ))\;
}

\caption{Strong Q-PUF: multi-key extraction from q-tuples}
\label{alg:strong-qpuf}
\end{algorithm}

The challenge with larger devices---such as IBM’s $127$-qubit quantum computers, which have been online since August $2024$---is that creating fingerprints by concatenating all qubit frequencies becomes impractical, as we explain here. The number of significant digits (in hexadecimal) used to store frequency data is fixed, so with this technique, the fingerprint size scales linearly with the number of qubits. As the fingerprint becomes longer, it becomes problematic for two key reasons. First, a large fingerprint results in a significantly large key, and sharing such a large key may not be feasible. Second, larger keys require a higher error threshold, as they are more susceptible to errors, and the implementation of a fuzzy extractor scales poorly as error tolerance increases.

To address the aforementioned issue, we propose using a subset of qubits from larger quantum devices, rather than utilizing all available qubits. Specifically, we introduce a new parameter, \textit{q-tuple}, defined as a subset of qubits whose properties are employed to generate the fingerprint. The fingerprint is constructed by concatenating the frequency data of the qubits in the q-tuple. Figure~\ref{fig:q_tuple} illustrates an example of q-tuple, shown as circled qubit on the topology of an IBM $127$-qubit device. Importantly, a q-tuple can take on $n^k$ distinct values, where $n$ is the total number of qubits in the quantum computer and $k$ is the size of the q-tuple. The exponential number of possible q-tuples makes this approach well-suited as a challenge mechanism for both \textit{weak} and \textit{strong} Q-PUF. 

\subsubsection{Weak and Strong Q-PUF Protocols}
Building on the frequency-based Q-PUF key extraction described earlier, we extend the protocol to support both \textit{weak} and \textit{strong} Q-PUF constructions using the concept of a \textit{q-tuple}—a selected subset of qubits from the device. This enables scalability to larger quantum systems while unlocking a richer challenge-response structure.

Protocol~\ref{alg:qtuple-qpuf} shows the \textit{weak} Q-PUF variant, where a fixed q-tuple is selected to generate a single key. This approach improves scalability by limiting the fingerprint size while still leveraging device-unique properties.

In contrast, protocol~\ref{alg:strong-qpuf} implements the \textit{strong} Q-PUF variant by generating multiple q-tuples, each of which produces a distinct key. This supports the generation of an exponential number of challenge-response pairs (CRPs), satisfying the design criteria of \textit{strong} PUFs. In this setting, different q-tuples serve as challenges, and the corresponding reproducible keys form the responses, all derived from the same quantum device.
\black

\subsubsection{Quality analysis of keys generated from q-tuples on large devices}

\begin{figure}[t]
     \centering
         \includegraphics[width=0.95\linewidth,trim={0cm 0cm 0cm 0cm},clip]{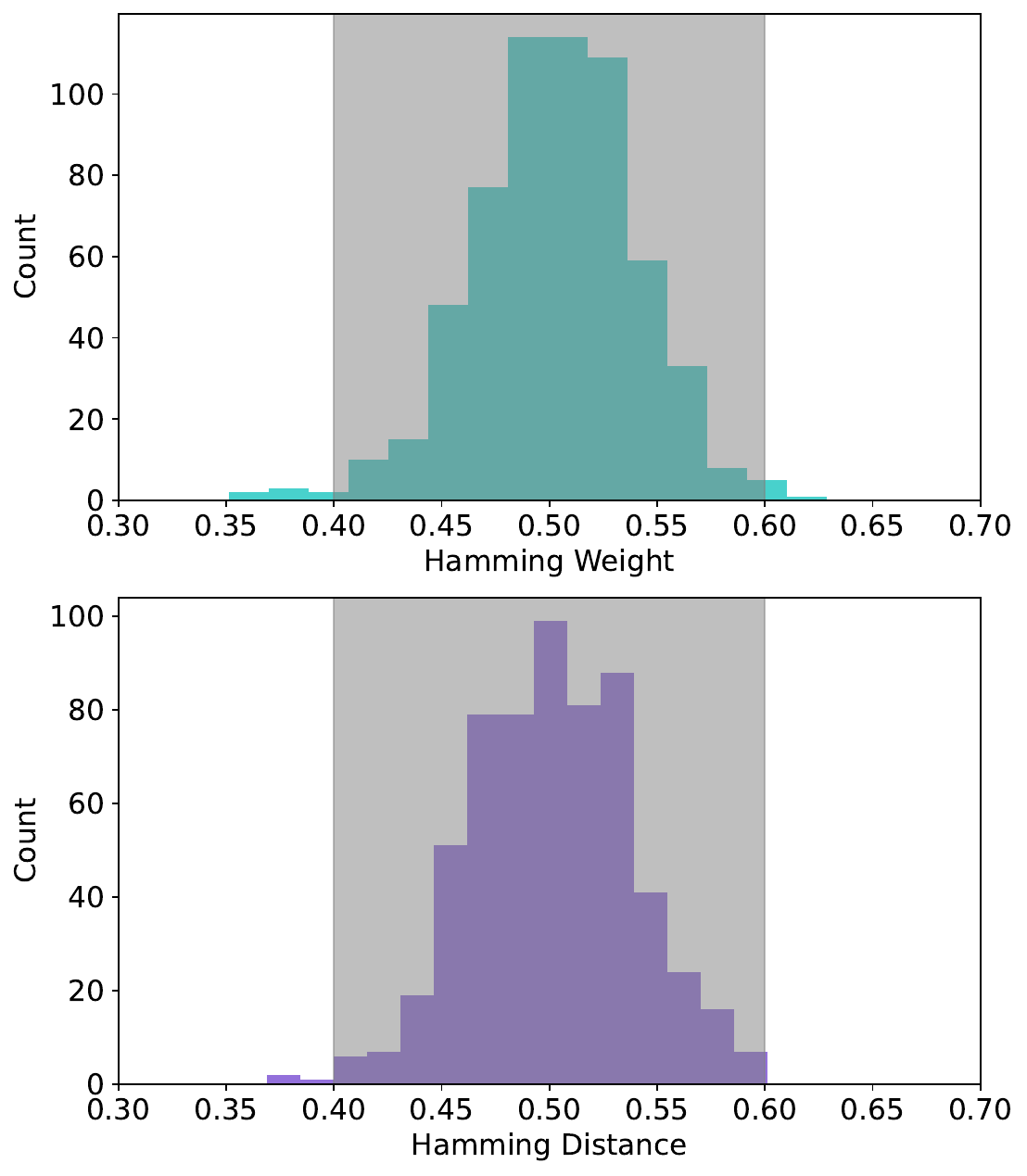}
        \caption{Hamming analysis of QPUF key generated on $3$ IBM 127-qubit devices using $100$ different choices of q-tuples. (300 data in total). 
        }
        \label{fig:hamming-analysis-large}
\end{figure}

We assess the quality of the Q-PUF protocol for larger devices, using the same methodology as the previous section---specifically evaluating the Hamming weight and Hamming distance of the generated keys. For this analysis, we used frequency data from three IBM $127$-qubit devices. From each device, we randomly selected $100$ distinct q-tuples of size $7$, resulting a total of $300$ datasets, each containing $7$ elements (qubit frequencies).

The results of the Hamming analysis are shown in Fig.~\ref{fig:hamming-analysis-large}. Both the Hamming distance and Hamming weight distributions exhibit near-Gaussian behavior centered around a mean of $0.5$, indicating that the keys have high entropy and are uniformly distributed.

\subsection{Using Anharmonicity as a Fingerprint}
One way to reduce the risk of fingerprint compromise is to derive it from multiple physical properties. For example, in superconducting fixed-frequency qubits used in IBM devices, we observe that---besides qubit frequency---anharmonicity can also serve as a viable fingerprinting feature.

\begin{figure}[t]
     \centering
     \includegraphics[width=0.95\linewidth,trim={.2cm 0cm 0cm 0cm},clip]{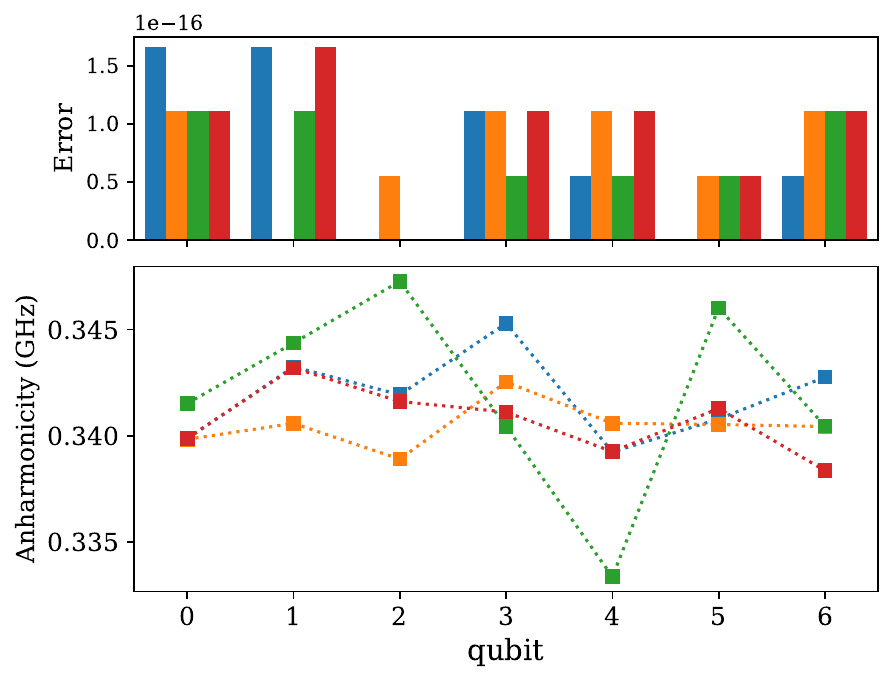}
        \caption{Average anharmonicity of the qubits in IBM's 7-qubit devices, along with their corresponding errors (refer to Fig.~\ref{fig:7q-T1-freq} for the color legend). The anharmonicity error for some qubits is below $10^{-16}$, making it too small to be shown in the error plot, and thus appears missing. }
        \label{fig:7q-anharm}
\end{figure}

Figure~\ref{fig:7q-anharm} shows the mean anharmonicity values of qubits in IBM’s 7-qubit devices, along with corresponding error bars. The measurement errors are on the order of $10^{-16}$, suggesting high stability and suitability for fingerprint generation. Our analyses confirm that anharmonicity-based fingerprints perform comparably to those based on frequency.

This opens the possibility for device owners to choose between frequency, anharmonicity, or a combination of both when generating fingerprints—without revealing which property is used. A more detailed study of this hybrid fingerprinting strategy, and its effectiveness as the basis for a Q-PUF, is left for future work.

\black

\section{Future Directions}

In this section, we first discuss how our Q-PUF proposal can be generalized to alternative technologies, with a specific emphasis on cold-atom quantum devices. We then highlight considerations for adapting the Q-PUF to utility-scale quantum systems. We leave the full development of these frameworks for future work.

\subsection{Extensions to Cold-Atom Technologies}

Our Q-PUF protocol focuses on transmon qubit technologies; however, the techniques can be extended to cold-atom platforms.  Cold-atom-based sensors, clocks, and computing hardware are poised to become critical elements of the future quantum computing ecosystem. Securing these end nodes is essential, especially as they may be responsible for delivering mission-critical data in potentially insecure environments. 

Similar to transmon-based QCs, we seek intrinsic properties that can serve as reliable and robust fingerprints, forming the basis for a Q-PUF. However, fingerprinting cold-atom devices requires fundamentally different methods due to the unique nature of these devices. For instance, a defining feature of the neutral-atom (NA) quantum computing platforms is that their qubits--- individual atoms suspended by optical tweezers---are physically identical. While this identical nature offers significant technical advantages for controlling the qubits, it prevents the use of intrinsic features for device fingerprinting, unlike in transmon-based QCs. This limitation motivates us to explore other aspects of the NA system architecture in order to develop suitable fingerprints and corresponding Q-PUFs.

However, cold-atom quantum devices exhibit systematic imperfections that can serve as the basis for an intrinsic Q-PUF fingerprint. In the remainder of this subsection, we identify components of cold-atom systems that offer viable sources of fingerprinting features and outline methods for their characterization. A detailed evaluation of the uniqueness, consistency, and collision resistance of these proposed cold-atom fingerprints is left for future work. 

\subsubsection{Control Hardware}
Neutral-atom quantum systems rely on sophisticated laser networks for atom cooling, trapping, addressing, and readout. Although laser sources must be ultra-precise, the control hardware is subject to operational variations~\cite{wintersperger2023neutral}. In particular, laser intensity and phase noise have been observed. Intensity noise---fluctuations in the beam's optical power---can be characterized using a laser spectrum analyzer, while phase noise---variation in optical phase---can be analyzed in the frequency domain using a standard spectrum analyzer.

\subsubsection{Vapor Cell}

Neutral-atom QCs and quantum sensors based on Rydberg atoms incorporate a glass cell as a core component, where atoms are trapped in a vacuum for quantum operations. Although these glass cells are manufactured to high precision, slight variation can arise due to fabrication limits. Certain noise channels, such as measurement uncertainty with Rydberg atom RF sensors, are influenced by the dimensions and properties of the vapor cell~\cite{holloway2015atom}. For instance, the RF field can interact with the glass, producing standing waves from internal reflection. Characterizing these standing waves may enable device-specific fingerprinting of cold-atom quantum hardware.

Moreover, fabrication imperfections can cause warping in vapor cell windows, particularly near the edges~\cite{ma2020dc}. Rydberg EIT spectroscopy can detect absorption changes caused by this warping, offering another potential avenue for fingerprinting.

\subsection{Scaling the Q-PUF}

Our efforts to establish trustworthy quantum computation through Q-PUFs has extensions divided into two fronts. First, we plan to expand the technology readiness of the physical Q-PUF based on intrinsic device fingerprints ensure its deployment whenever hybrid networks that include practical quantum hardware are on the horizon. Tasks in this category include developing a Q-PUF for alternate qubit technologies, such as cold-atoms, 
and exploring candidate physical implementations for the fuzzy extractor logic and challenge-response framework. Second, we aim to extend Q-PUF concepts to logical qubits.

\subsubsection{Expanding the Q-PUF Extractor Function}

In this initial demonstration of our Q-PUF, we present a software prototype of the fuzzy extractor that generates a secure key from the frequency-based fingerprint. In the next steps of this work, we plan to expand our framework to explore candidate designs for the fuzzy extractor's hardware architecture. 
An open-ended question that we aim to solve is discovering the optimal location for a crypto-logic peripheral relative to the quantum device. 

\subsubsection{Logical Q-PUFs}

\begin{figure}
\centering
\includegraphics[width=\columnwidth,trim={0 7.5cm 0 7.5cm},clip]{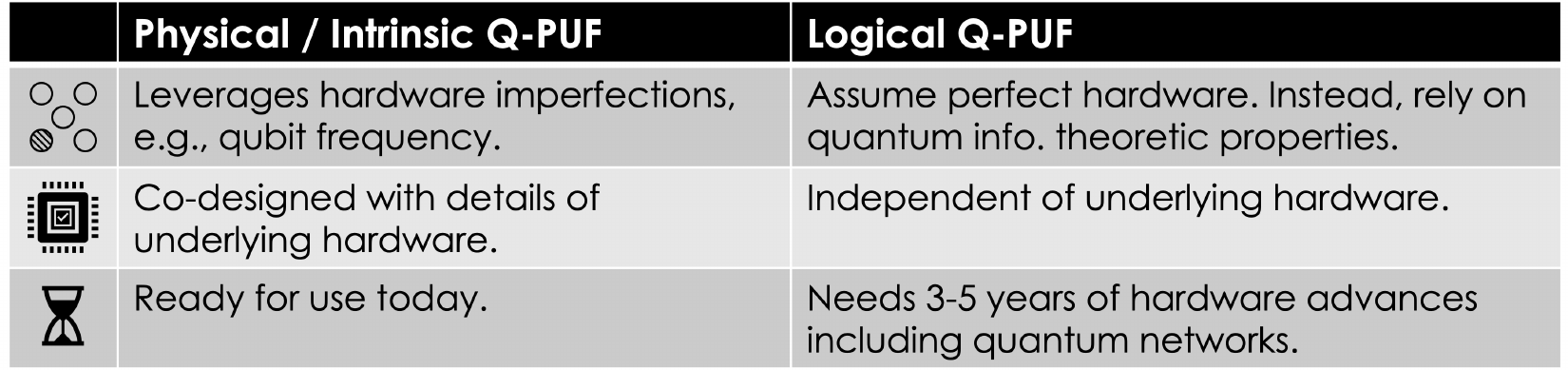}
\caption{This work primarily focused on Physical / Intrinsic Q-PUFs. In future work, we aim to develop Logical Q-PUFs. The distinction is similar to the relationship between physical and logical qubits.}
\label{fig:physical_logical_qpuf}
\end{figure}

Our discussion so far has covered efforts towards developing Q-PUFs that rely on physical imperfections in quantum hardware. In anticipation of the fault-tolerant (FT) quantum systems of the future, we are motivated to develop an alternative, and complementary, approach to Q-PUFs that relies on more fundamental information theoretic properties of quantum mechanics. As shown in Fig.~\ref{fig:physical_logical_qpuf}, this bifurcation of Q-PUFs mirrors the distinction between physical and logical qubits. As such, we refer to this future work as logical Q-PUFs. With a logical Q-PUF, a QC user could not only confirm the identity of the origin of their results, they would also be able to certify that the hardware satisfies DiVincenzo's Criteria for quantum computation~\cite{divincenzo2000physical}.

\section{Related Work}

Prior work has proposed PUFs targeted for quantum computing applications. For example, a classical PUF based on the quantum readout of unclonable quantum states is found in~\cite{vskoric2012quantum}. Second, the work in~\cite{phalak2021quantum} proposed quantum hardware PUFs that generate bit strings with functions that apply qubit superposition and decoherence. Next, the work in~\cite{arapinis2021quantum} and~\cite{doosti2021client} present theory for quantum PUFs that establish a secure communication channel using quantum properties and operations that are secure against quantum cryptographic attacks. There is prior art in the space of quantum device identification based on crosstalk~\cite{allen2021short} and the diversity and instability of quantum errors~\cite{wu2024detecting}. Finally, proposals for quantum antivirus exist~\cite{deshpande2022towards} that aim to flag suspicious programs that inject malicious crosstalk and degrade the quality of program outcomes. Our work differs from these past efforts by expanding on the methods of~\cite{smith2023fast} to combine a frequency-based fingerprint with fuzzy extraction to create a Q-PUF. 

\section{Conclusion}
Remotely accessible quantum computers represent a critical shift in how quantum computation is performed—but they also introduce new security risks. In this work, we proposed and demonstrated a Q-PUF scheme that uses intrinsic quantum hardware features, specifically qubit frequencies, in combination with fuzzy extractors to enable device authentication and secure key generation. We validated the effectiveness of our protocol using real IBM QCs and large-scale simulations, showing that consistent and unique keys can be reliably extracted.

While our initial implementation represents a weak PUF, we showed how the method can be extended to large-scale quantum systems using q-tuples, enabling scalable key generation and supporting a rich challenge-response architecture. This lays the foundation for strong Q-PUF constructions, a key step toward practical, secure quantum hardware authentication in the cloud.

\section{Acknowledgments}
This material was supported by the Australian Army and was featured in the Australian Army's Quantum Technology Challenge (QTC) to prevent disruption of QCs and ensure their reliability and resilience. We acknowledge the use of IBM Quantum services for this work. The views expressed are those of the authors, and do not reflect the official policy or position of IBM or the IBM Quantum team.

\bibliographystyle{IEEEtran}
\bibliography{references}

\end{document}